\documentclass[preprint]{aastex61}

\def\kms{$\rm{km~s}^{-1}$}

\newcommand{\hii}{\hbox{H\,{\sc ii}}}
\newcommand{\ha}{\hbox{H$\alpha$}}
\newcommand{\hb}{\hbox{H$\beta$}}
\newcommand{\hd}{\hbox{H$\delta$}}
\newcommand{\hg}{\hbox{H$\gamma$}}

\newcommand{\oi}{\hbox{[O\,{\sc i}]}}
\newcommand{\oii}{\hbox{[O\,{\sc ii}]}}
\newcommand{\nii}{\hbox{[N\,{\sc ii}]}}
\newcommand{\sii}{\hbox{[S\,{\sc ii}]}}
\newcommand{\oiii}{\hbox{[O\,{\sc iii}]}}
\newcommand{\neiii}{\hbox{[Ne\,{\sc iii}]}}
\newcommand{\hei}{\hbox{[He\,{\sc i}]}}

\newcommand{\zoh}{\hbox{$12\,+\,{\rm log(O/H)}$}}

\newcommand{\Msun}{\hbox{$M_{\sun}$}}
\newcommand{\Mstar}{\hbox{$M_{*}$}}
\newcommand{\te}{\hbox{$T_e$}}

\received{*******}
\revised{*******}
\accepted{********}
\submitjournal{***}

\shortauthors{Hu et al.}

\begin{document}

\title{M101: Spectral observations of \hii\ regions and their physical properties}

\author{Ning Hu}
\affiliation{CAS Key Laboratory for Research in Galaxies and Cosmology, Department of Astronomy, University of Science and Technology of China, Hefei 230026, China}
\affiliation{School of Astronomy and Space Sciences, University of Science and Technology of China, Hefei 230026, China}
\email{huning@mail.ustc.edu.cn}

\author{Enci Wang}
\affiliation{CAS Key Laboratory for Research in Galaxies and Cosmology, Department of Astronomy, University of Science and Technology of China, Hefei 230026, China}
\affiliation{School of Astronomy and Space Sciences, University of Science and Technology of China, Hefei 230026, China}

\author{Zesen Lin}
\affiliation{CAS Key Laboratory for Research in Galaxies and Cosmology, Department of Astronomy, University of Science and Technology of China, Hefei 230026, China}
\affiliation{School of Astronomy and Space Sciences, University of Science and Technology of China, Hefei 230026, China}

\author{Xu Kong}
\affiliation{CAS Key Laboratory for Research in Galaxies and Cosmology, Department of Astronomy, University of Science and Technology of China, Hefei 230026, China}
\affiliation{School of Astronomy and Space Sciences, University of Science and Technology of China, Hefei 230026, China}

\author{Fuzhen Cheng}
\affiliation{CAS Key Laboratory for Research in Galaxies and Cosmology, Department of Astronomy, University of Science and Technology of China, Hefei 230026, China}
\affiliation{School of Astronomy and Space Sciences, University of Science and Technology of China, Hefei 230026, China}

\author{Zou Fan}
\affiliation{Key Laboratory of Optical Astronomy, National Astronomical Observatories, Chinese Academy of Sciences, 20A Datun Road, Chaoyang District, Beijing 100012, China}

\author{Guanwen Fang}
\affiliation{Institute for Astronomy and History of Science and Technology, Dali University, Dali 671003, China}

\author{Lin Lin}
\affiliation{Shanghai Astronomical Observatory, Chinese Academy of Science, 80 Nandan Road, Shanghai, 200030, China}

\author{Yewei Mao}
\affiliation{Purple Mountain Observatory, Chinese Academy of Sciences, Nanjing 210008, China}

\author{Jing Wang}
\affiliation{CSIRO Astronomy \& Space Science, Australia Telescope National Facility, P.O. Box 76, Epping, NSW 1710, Australia}

\author{Xu Zhou}
\affiliation{Key Laboratory of Optical Astronomy, National Astronomical Observatories, Chinese Academy of Sciences, Beijing 100012, China}

\author{Zhiming Zhou}
\affiliation{Key Laboratory of Optical Astronomy, National Astronomical Observatories, Chinese Academy of Sciences, Beijing 100012, China}

\author{Yinan Zhu}
\affiliation{Key Laboratory of Optical Astronomy, National Astronomical Observatories, Chinese Academy of Sciences, 20A Datun Road, Chaoyang District, Beijing 100012, China}

\author{Hu Zou}
\affiliation{Key Laboratory of Optical Astronomy, National Astronomical Observatories, Chinese Academy of Sciences, Beijing 100012, China}

\begin{abstract}

By using the Hectospec 6.5 m Multiple Mirror Telescope (MMT) and the 2.16 m telescope of National 
Astronomical Observatories, Chinese Academy of Sciences (NAOC), we 
obtained 188 high signal-to-noise ratio (S/N) spectra of \hii\ regions in the nearby galaxy M101,
which are the largest spectroscopic sample of \hii\ regions for this galaxy so far.
These spectra cover a wide range of regions on M101, which enables us 
to analyze two dimensional distributions of its physical properties. 
The physical parameters are derived from emission lines or stellar continuum, including
 stellar population age, electron temperature, oxygen abundance and etc.
The oxygen abundances are derived using two empirical methods based on O3N2
 and R$_{23}$ indicators, as well as the direct \te\ method when $\oiii\lambda4363$ 
 is available. 
By applying the harmonic decomposition analysis to the velocity field, we 
obtained line-of-sight rotation velocity of 71 \kms\ and a position angle of 36 degree. 
The stellar age profile shows an old stellar population in galaxy center and a relative young 
stellar population in outer regions, suggesting an old bulge and a young disk. 
Oxygen abundance profile exhibits a clear break at $\sim$18 kpc, 
with a gradient of $-$0.0364 dex kpc$^{-1}$ in the inner region and $-$0.00686 dex kpc$^{-1}$
 in the outer region. Our results agree with  the  ``inside-out'' disk growth scenario of M101.
\end{abstract}


\section{Introduction}

$\hii$ regions are the sites of strong star formation in galaxies, which make \hii\ regions perfect
probes of star formation processes, evolution of massive stars and the surrounding interstellar medium.
Plenty of information can be obtained by analyzing the emission lines and underlying stellar continuum
of their spectra. The gas-phase metallicity, defined as the number ratio of oxygen to hydrogen atom,
 is one of the key properties in galaxy formation and evolution. 
 The oxygen is synthesized in high-mass stars (\textgreater8\Msun) and then released
to the inter-stellar medium through stellar winds or supernova explosion. 
The spatial distributions of the oxygen abundance in galaxies are affected by a variety of processes, 
such as enriched outflows \citep{tre04}, accretion \citep{dal04},
and mergers \citep{kew06,rup10,kew10,rich12,torr12,san14}. Thus studying the metallicity
and its relation with other properties would provide clues on the galaxy formation and evolution.

Researches on the metallicity of \hii\ regions are booming in the last two decades. 
\citet{zaritsky94} and \citet{zee98} found that \hii\ regions have an 
average metallicity gradient of $-$0.05 dex kpc$^{-1}$, i.e. the inner regions of galaxies
are more metal-rich than the outskirts. The negative gradients are found to be universal, 
as reviewed by many following researches \citep{bre07,scar13,li13,san14,ho15,san16b}, 
suggesting inner-to-out transportation of metals. 
Observations also show breaks in the oxygen abundance gradients in a number 
of galaxies \citep{zaritsky92,roy97}. \citet{pil03} claimed that such breaks 
are due to the systematic error involving the excitation parameter, while others attribute it 
to the barrier effect of corotation, which isolates the inner and outer regions of the disk one
 from the other due to opposite directions of gas flow \citep{lep11,scar13}.
More recently, several studies focusing on gas content have found that the metallicity
gradients flatten to a constant value beyond the isophotal radii R$_{25}$ or 2R$_e$\ \citep{ros11,mar12,
san12b,san14,lop15,san16b}, and several scenarios are proposed to explain the nature of this flattening,
such as the bar induced radial gas flows \citep{cav14}, minor mergers and perturbations from satellite
galaxies \citep{bird12,lop15}, the varying star formation efficiency over
a large galacentric distances \citep{bre12,est13}, and the balance between outflows and inflows with
the intergalactic medium \citep{opp08,opp10,dav11,dav12}.

Recently, the Integral Field Unit (IFU) spectrograph is becoming a 
most important and powerful tool for spatial-resolved spectroscopic observations, which 
has greatly increased the progress of the two-dimensional research on galaxies.
Many surveys using IFU technology are carried out, such as SAURON \citep{bac01},  
PINGS \citep{ros10},  CALIFA \citep{san12a} and MaNGA \citep{bun15} survey. 
According to their different observational strategies, MaNGA is mapping a large 
number of galaxies with limited spatial resolution (1-3 kpc) and field of view ($\sim$70 acrsec),
while PIGNS provides a better spatial resolution and larger field of view but with limited 
number of galaxies, and CALIFA is a compromise between the two. 
\citet{san14} found a common metallicity
gradient of $-$0.10$\pm$0.09 dex R$_e^{-1}$\ for 193 CALIFA galaxies, and the metallicity gradients 
do not exhibit the dependence on other properties of galaxies, such as morphology, 
mass, and the presence or absence of bars \citep{san16b}. 
By studying 49 local field star-forming galaxies, \citet{ho15} proposed a local benchmark of 
metallicity gradients of $-$0.39$\pm$0.18 dex R$_{25}^{-1}$. 
More recently, \citet{bel17}  analyzed 550 nearby galaxies from MaNGA survey and 
confirm that metallicity gradient is flat for low mass galaxies ($\Mstar<10^{9.0}\Msun$), 
steepens for more massive galaxies until $\Mstar\sim10^{10.5}\Msun$\ 
and flattens lightly again for even more massive galaxies.
These researches provided strong constraints on galactic chemical composition of nearby 
galaxies \citep{chi01,fu09,ho15,san16b,bel17}. 

\citet{kong14} proposed  the ``Spectroscopic 
Observations of the \hii\ Regions In Nearby Galaxies (H2ING)" Project, which performs spectroscopic 
observations on \hii\ regions in 20 nearby large galaxies since 2008. As the third paper of this project,
we report the spectroscopic observation 
of \hii\ regions in nearby galaxy M101, with the MMT 6.5 m telescope \citep{fab05} and the NAOC 2.16 m 
telescope \citep{2.16}. M101 (also known as NGC 5457, $\alpha=14^h03^m12^s.5$, 
$\delta=+54^\circ20'56''$) is a large face-on Scd galaxy containing plenty of \hii\ regions \citep{hod90, 
 kenn03, gor08, bre07, cro16}. 
It has a distance of about 7.4 Mpc and an apparent scale of about 36 pc arcsec$^{-1}$, 
which enables us to observe its \hii\ regions to a scale of a few hundred pc with our
 $1.5''$\ fiber and $2.5''$\ slit observations.
A major-axis position angle of $39^\circ$ and an inclination angle 
of $18^\circ$ \citep{bosma81} allow us to perform detailed studies of 
its stellar populations and ionized nebulae.
Observations of the M101 \hii\ regions have been carried out since 1970s \citep{sea71,smith75,mc85,kenn96,bre07}. 
Recently, \citet{li13} presented a catalog containing 79 \hii\ regions from their observations 
and several previous works since 1996. 
\citet{cro16} have enlarged the sample of \hii\ regions by using the Large Binocular Telescope (LBT), 
and obtained 109 spectra of \hii\ regions in M101. 
In this paper, we have obtained 188 \hii\ region with high S/N from our observations, 
which is the largest spectroscopic sample of \hii\ regions for M101. 

With the spectra of numerous \hii\ regions, we have derived the physical properties of \hii\ regions, 
and presented a detailed study of the spatial distribution of these properties.
This paper is structured as follows. In Section 2, we describe the observation, 
data reduction and emission line measurements. 
In Section 3, we present the measurements and analysis of physical properties of the \hii\ region sample.
Metallicities are derived with three methods and the gradient is also calculated and discussed. 
We summarize our results in Section 4.
 
\section{Data}
\subsection{Observations}

The \hii\ region candidates are selected from the continuum-subtracted \ha\ image 
\citep{hoo01}. Candidates are primarily selected with preference for large regions, using SExtractor 
\citep{ber96} with irregular area filter with at least 25-pixel (1 pixel $=2.028''$).
The foreground stars are excluded from the \hii\ catalog by matching with
 the 2MASS all-sky Point Source Catalog \citep[PSC;][]{cut03}.
Then candidates are observed with the Hectospec multi-fiber positioner and spectrograph 
on the 6.5 m MMT telescope and the NAOC 2.16 m telescope.

The usable field of the MMT 6.5 m telescope is 1\arcdeg\ in diameter, and the instrument deployed 
three hundred $1.5''$-diameter fibers on the field, corresponding to $\sim$ 
54 pc at the distance of M101. We used Hectospecs 
270 gpm grating that provided a dispersion of 1.2 \AA\ pixel$^{-1}$ and 
a resolution of  $\sim$ 5\ \AA. The spectral wavelength coverage is from 3650 to 
9200 \AA.  The Hectospec fiber assignment software
\emph{xfitfibs}\footnote{https://www.cfa.harvard.edu/john/xfitfibs} allows the user to assign rankings 
to targets. If we ignored the spatial positions
of our targets, fiber collisions would prevent many objects in the center portion of the center of
M101 from being observed. We therefore assigned priority to \hii\ regions based on their \ha\ fluxes.
We observed two fields at the night on 2012 February 10 with
3600\ s exposure times and one field on 2013 March 15 with 5400\ s exposure times. Weather conditions
are clear on both nights, and seeings are about $1.2''$, $0.8''$ and $0.6''$ for each field, respectively.

The NAOC 2.16 m telescope worked with an OMR (Optomechanics Research Inc.) spectrograph providing 
a dispersion of 4.8\ \AA\ pixel\ $^{-1}$ and a resolution of 10\ \AA. The observed spectra cover 
the wavelength range of 3500 to 8100\ \AA.
Slits of $2.5''$ are placed manually to cover as many candidates as possible and to avoid those observed by
MMT. Observations are carried out over 9 nights between 2012 and 2014 with 20 slits, 
and exposure times varied between 3600\ s and 5400\ s depending on weather conditions. 
Typical seeing is $\sim 2.5''$, corresponding to an spatial resolution of $\sim$300 pc.   

\subsection{Data reduction}
\subsubsection{MMT spectra}
We obtained over 300 spectra from MMT observations, and these spectra are reduced 
in an uniform manner with the publicly available HSRED\footnote{http://mmto.org/~rcool/hsred/index.html} 
software. The frames are first de-biased and flat-fielded. Individual spectra are then extracted and wavelength calibrated. 
Sky subtraction is achieved with Hectospec by averaging spectra from ``blank sky'' fibers from the same exposures. 
Three spectra of the same target are reduced  and combined into one final spectrum. Standard stars are obtained 
intermittently and are used to calibrate spectra using IRAF ONEDSPEC package. These relative flux corrections 
are carefully applied to ensure that the relative line flux ratios are accurate. 
We have checked the spectra by visually inspection to exclude spectra with problematic continuum shape,
or missing \ha\ and \hb\ emission lines.
Finally, we obtained a sample of 164 \hii\ region spectra for analysis. 
The 164 spectra have a median S/N of 35 at 5000\AA, and over 50 spectra have S/N higher than 60.

\subsubsection{NAOC 2.16 m Spectra}
The spectroscopic data from NAOC 2.16 m observations are reduced following standard procedures using IRAF 
software package. The CCD reduction includes
bias and flat-field correction, as well as cosmic-ray removal. Wavelength calibration is performed 
based on helium/argon lamps exposed at both the beginning and the end of each slit during the observation. 
Flux calibration is performed based on observations of the KPNO spectral 
standard stars \citep{mas88} observed at the end of each slit observation. 
The atmospheric extinction is corrected by using the mean extinction coefficients measured for 
Xinglong by the Beijing-Arizona-Taiwan-Connecticut 
(BATC) multicolor sky survey (H. J. Yan 1995, priv. comm.). Similar to MMT spectra, 
those with problematic continuum shape or missing \ha\ and \hb\ emission lines are excluded.
 Finally we extracted 41 spectra, among which
17 spectra are also observed by MMT 6.5 m telescope. Considering the higher S/Ns and better resolution
of MMT spectra, we use MMT spectra for these 17 candidates, and NAOC observations provide 24 
additional spectra.

\begin{figure} \centering
\includegraphics[width=0.9\columnwidth]{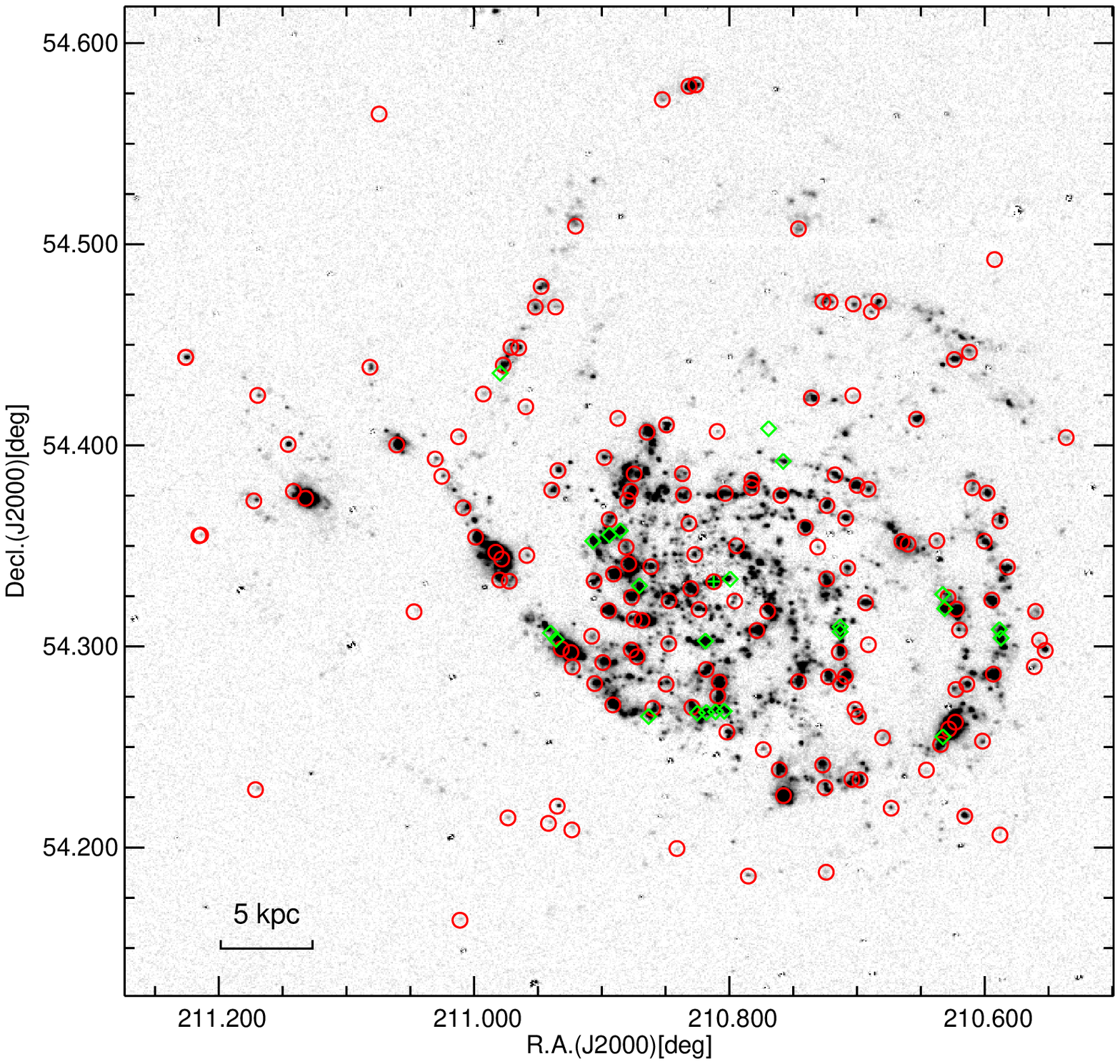}
\caption{ Location of \hii\ regions of final 188 samples on the narrow-band \ha\ image of M101 
observed with KPNO Schmidt telescope \citep{hoo01}. Two symbols represent spectra from two telescop:
red circles from MMT and green diamonds from NAOC. The green cross marks the center of the galaxy.}
\label{image}
\end{figure}

Finally we have 188 spectra in total (164 from MMT 6.5 m and 24 from NAOC 2.16 m). Locations of these \hii\ regions
 are shown in Figure\,\ref{image} and their coordinates are listed in Table\,\ref{data}.
Velocities are measured using the SAO xcsao package with IRAF (see Figure\,\ref{2d}). 
We have carefully checked the velocities and corrected a few bad results manually.
Spectra are corrected for Galactic reddening \citep{schlegel98}, and then shifted to 
rest frame for further measurements. 

\subsection{Spectral Fitting}

\begin{figure} \centering
\includegraphics[width=0.9\columnwidth,angle=270]{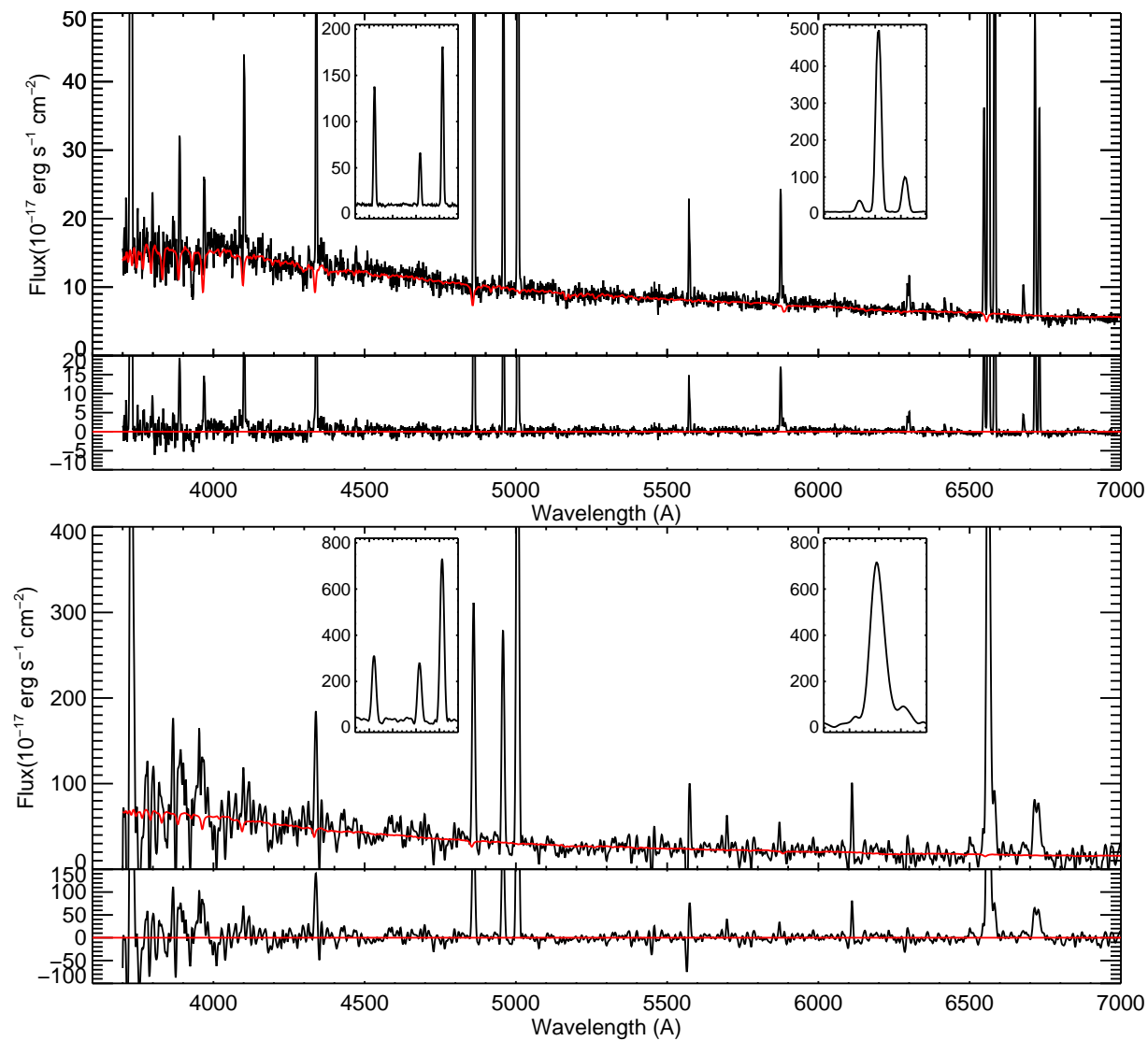}
\caption{Typical spectrum of \hii\ regions in M101 observed by MMT (upper panel) and NAOC 2.16m 
(lower panel). Panels show the observed spectra (black line) and
MF-ICA continuum fit (red line), and the two small insets of each spectrum zoom in several strong emission 
lines: \hb + \oiii$\lambda\lambda$4959,5007, \ha + \nii$\lambda\lambda$6548,6583.}
\label{sample}
\end{figure}

To measure the emission lines of each individual spectra, the underlying stellar continuum 
must be subtracted. We perform a modeling of stellar continuum using advanced ICA algorithm,
mean field approach to Bayesian independent component analysis (MF-ICA), which 
is comparatively precise and efficient \citep{hs01, hs02}. 
The MF-ICA approach extracts 12 independent components (ICs) from BC03\footnote{http://www.bruzual.org} \citep{bc03}. 
The 12 ICs contain full information of BC03 library and excellently recover the library, which 
spans a stellar age range from $1.0\times10^5$\ to $2.0\times10^{10}$\ yr,
 and an initial chemical composition metallicity from $0.0001$
to $0.1$. The star formation histories (SFHs) are parameterized in terms of an 
underlying continuous model superimposed with random bursts on it \citep{kau03}. 
The intrinsic starlight reddening is modeled by the extinction law of \citet{char00}. 
The velocity dispersion is set to vary between 50 $\rm{km\,s^{-1}}$\ and 
450 $\rm{km\,s^{-1}}$. MF-ICA provides reliable modeling of stellar continuum 
and estimates physical parameters accurately with a large improvement in efficiency. 
More detailed informations about MF-ICA algorithm are described in 
\citet{hu16}. Figure\,\ref{sample} shows two typical fitting results. For each spectrum, 
 emission lines $\oii\lambda3727$,  \hb + \oiii$\lambda\lambda$4959,5007, \ha + \nii$\lambda\lambda$6548,6583\
  and $\sii\lambda\lambda6717,6731$\ are enlarged in four small panels below the spectrum. 

\subsection{Line Flux Measurement}

After subtracting the stellar continuum, we perform the non-linear least-squares fit to emission lines 
using MPFIT package implemented in IDL \citep{mark09}. Each emission line is modeled with one
Gaussian profile, and we constrain the ratios \nii$\lambda$6583/\nii$\lambda$6548 and 
\oiii$\lambda$4959/\oiii$\lambda$5007 to their theoretical values given by quantum mechanics. 
 The emission line fluxes are measured by integration of the flux based on the fitted profiles. 

The interstellar reddening is corrected by comparing the observed \ha/\hb\ Balmer decrement to the 
theoretical value, since intrinsic value of \ha/\hb\ is not very sensitive to the physical conditions
of the gas. Assuming $\ha/\hb=2.86$ and $R_V=3.1$ with \citet{card89} extinction curve under the
case-B recombination, the extinction in the V band is given by:
\begin{equation}\centering
A_V=6.77\log \frac{F(\ha)/F(\hb)}{2.86}.
\end{equation}
negative color excesses are all set to be zero.
The reddening-corrected line flux measurements are listed in Table\,\ref{data}, with values 
normalized to \hb\ flux. All spectra have strong \ha\ and \hb\ emission lines, and 14 of them
have \oiii$\lambda$4363 emission line. 

\begin{figure} \centering
\includegraphics[width=0.9\columnwidth]{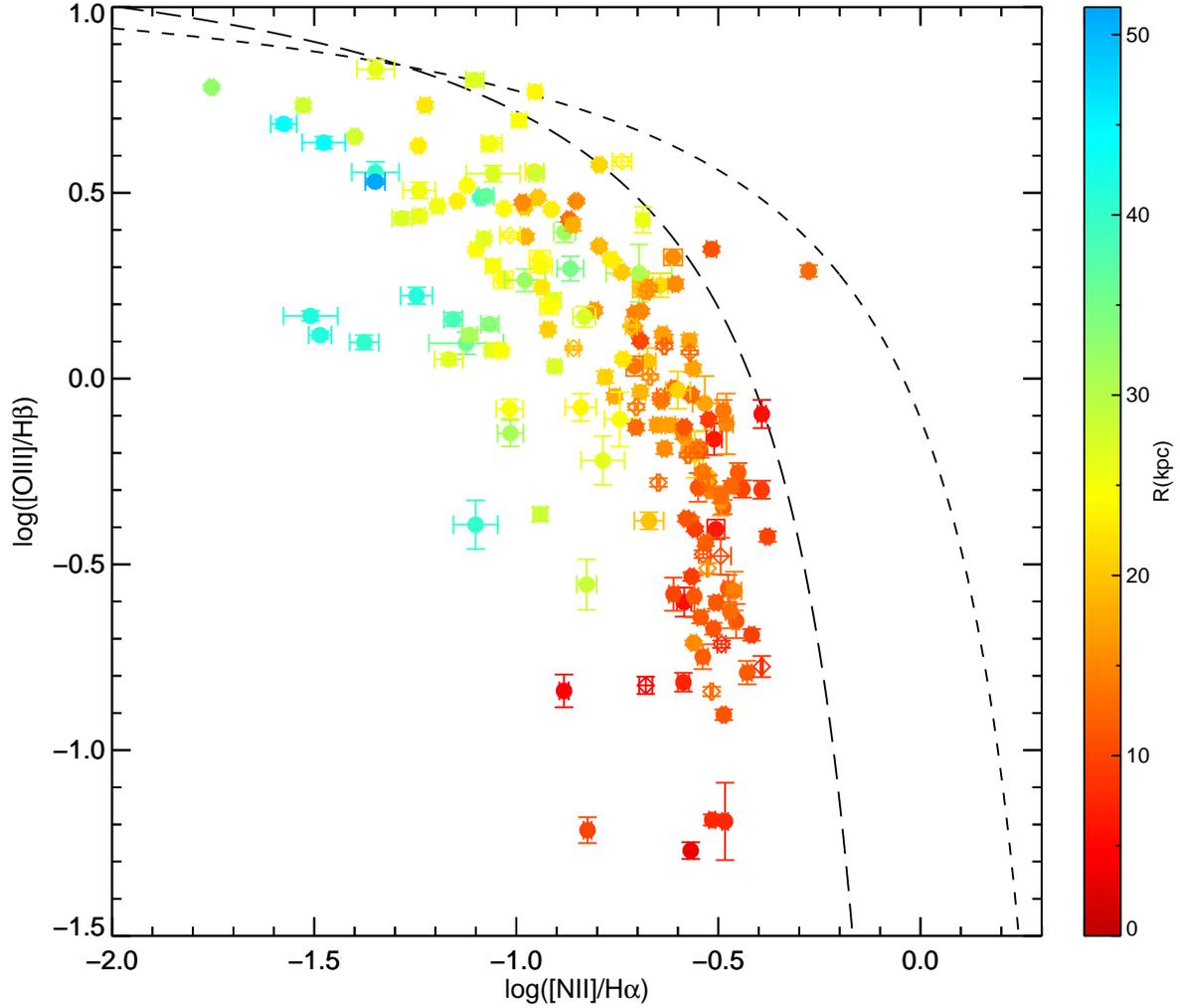}
\caption{BPT diagram showing the excitation properties of our \hii\ regions. The boundaries are
taken from \citet[short-dashed line]{kew01} and \citet[long-dashed line]{kau03}.
Filled points are from MMT and diamonds are from NAOC, and colors indicate the de-projected distance to 
the galaxy center. Almost all targets in our sample are located in the pure star forming region of the diagram with 
a clear profile against the de-projected distances to the galaxy center.
}
\label{figBPT}
\end{figure}

In Figure\,\ref{figBPT}, we examine the excitation properties of our sample by plotting
\oiii$\lambda$5007/\hb\ versus \nii$\lambda$6583/\ha\ diagnostic diagrams \citep[BPT,][]{BPT}.
The color-code indicates the de-projected distance to the galaxy center.
We plot the boundaries between different photoionization sources (star-forming
regions and active galactic nuclei) by \citet{kew01} and \citet{kau03}. 
As shown in Figure\,\ref{figBPT}, almost all targets in our sample are located in the 
pure star forming region with only 5 exceptions. 
Further, the \hii\ regions show a radially changing sequence on the BPT diagram: the inner regions have 
higher \nii$\lambda$6583/\ha\ and lower \oiii$\lambda$5007/\hb\ than outer regions, suggesting a
negative gradient of metallicity. This is well consistent with the results from \citet{san15},
 that the inner \hii\ regions of the galaxies appear to have higher \nii$\lambda$6583/\ha\ and lower 
\oiii$\lambda$5007/\hb\ than outer regions, using over 5000 \hii\ regions from 306 galaxies.

More detailed information will be discussed in section 3.3.

\subsection{Oxygen Abundance determination}
\subsubsection{The direct method}
The direct method to derive oxygen abundance is to measure the ratio of a higher excitation line to 
a lower excitation line, in which case $\oiii\lambda4363/\oiii\lambda\lambda4959,5007$ is the most commonly
used. This ratio together with a model of two excitation zone structure for the \hii\ regions provides an estimation of
the electron temperature of the gas, and then the electron temperatures are converted to oxygen abundances
with corrections for unseen stage of ionization. 

To calculate the total oxygen 
abundances, we make the usual assumption:\ $\rm{O/H} = (\rm{O}^++ \rm{O}^{2+})/\rm{H}^+$. 
Electron temperatures are derived based on the emission line intensity ratio $\oiii\lambda4363/\oiii\lambda\lambda4959,5007$, 
and electron densities ($n_{\rm{e}}$) are derived based on the intensity ratio $\sii\lambda6717/\sii\lambda6731$. 
Among the spectra of \hii\ regions, 14 of them have 
$\oiii\lambda4363$ detections with well-defined line profiles and the S/Ns larger than 3.  
The electron temperatures of them for the $\rm{O}^{2+}$-emitting region ($T\oiii$) are derived 
with the PYNEB package \citep{lur15}. 
PYNEB is a python package for analyzing emission lines, which includes the Fortran code FIVEL 
\citep{de87} and the IRAF nebular package \citep{shaw95}. We adopted the transition probabilities of 
\citet{wie96} and \citet{stor00} for $\rm{O}^{2+}$, \citet{podo09} for $\rm{S}^{+}$, the collision strengths of
\citet{agga99} for $\rm{O}^{2+}$, \citet{taya10} for $S^{+}$. Electron temperatures and electron
 densities are calculated simultaneously to obtain self-consistent estimates by PYNEB. 
Then the electron temperatures of \oii\ ($T\oii$) are obtained with applying the relation from \citet{g92}:   
$T\oii=0.7\times T\oiii+3000\rm{K}$. With giving $T\oii$, $T\oiii$\ and $n_{\rm{e}}$, 
metallicities are finally calculated from the line intensities of $\oii\lambda3727$\ and $\oiii\lambda\lambda4959,5007$\
with PYNEB.
 
Errors are estimated with Monte Carlo algorithm by repeating the calculation 2000 times. 
In each calculation, the input flux of each emission line is generated from a Gaussian distribution
with mean value equaling to measured emission line flux and standard deviation equaling to its error. 
We use the median value of 2000 results as the final value, and the half of $16\%-84\%$ range of this 
distribution as the corresponding error. The electron temperatures and densities are listed in Table \ref{metadata}.

Although the direct method is considered to be one of the most reliable methods 
to derive gas-phase oxygen abundances, the $\oiii\lambda4363$\ line is weak, 
especially in meta-rich environments. 
Since most of our \hii\ regions do not have $\oiii\lambda4363$ detections, we applied two 
strong-line methods to determine the oxygen abundance.

\subsubsection{O3N2 index}
The O3N2 index, defined as $\rm{O3N2}=\log\{(\oiii\lambda5007/\hb)/(\nii\lambda6583/\ha)\}$, 
is firstly introduced by Alloin et al. (1979). It is widely used to diagnose oxygen abundances in the literature. 
The four lines involved in the index are easily detected, and the close wavelength between the two pairs of lines 
make the index nearly free from extinction.
By using 137 extragalactic \hii\ regions, \citet{pett04} empirically calibrated the O3N2 vs oxygen abundance 
determined by \te\ method and photoionization models. More recently, \citet{m13} provide 
a new calibration with \te\ based metallicities of 603 \hii\ regions, which is adopted here:
\begin{equation}\centering
12+\log \rm{(O/H)}=8.533-0.214\times\rm{O3N2},
\end{equation}
However, O3N2 calibrations do not consider the variations of ionization parameter ($q$), which may cause 
systematic errors. More will be discussed in the next subsection. 

\subsubsection{KK04 method}
\citet[hereafter KK04]{kk04} adopted the stellar evolution and photoionization from \citet{kd02} to 
produce a modified $R_{23}$ calibration. $R_{23}$ index, defined as
$R_{23}=(\oii\lambda3727+\oiii\lambda\lambda4959,5007)/\hb$, 
 is sensitive to the ionization parameter of the gas, which is defined as the ratio between the 
number of hydrogen ionizing photons passing through a unit area per second and the hydrogen density 
of the gas. The ionization parameter is typically derived based on the line ratio
 $O_{32}=\oiii\lambda5007/\oii\lambda3727$, 
and it is also sensitive to oxygen abundance, therefore KK04 applied an iterative process to derive a consistent 
ionization parameter and oxygen abundance. First we determined whether the \hii\ regions lie on
the upper or lower $R_{23}$ branch with ratio $\rm{N2O2}=\nii\lambda6583/\oii\lambda3727$, then we calculated
an initial ionization parameter by assuming a starting oxygen abundance 8.0 for lower branch and 9.0 for
upper branch. The initial ionization parameter is used to calculate an initial oxygen abundance. These processes
are iterated until oxygen abundance converges. Errors are generated with the same Monte Carlo algorithm 
used in direct \te\ method, i.e. repeating the calculation 2000 times.

\begin{figure} \centering
\includegraphics[width=0.9\columnwidth]{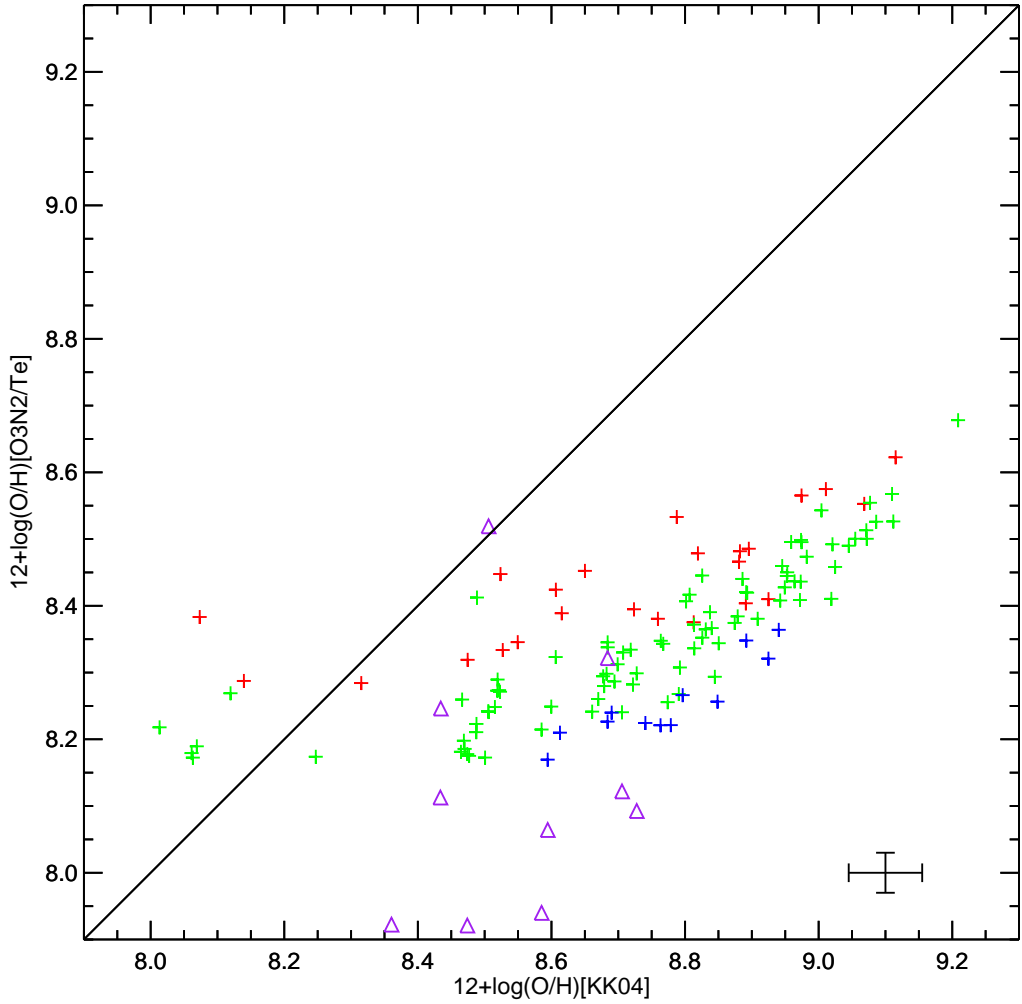}
 \caption{Comparation of the three abundance calibrations. Solid line is the y=x line. Triangle points represent 
direct \te\ calibration, and cross points represent O3N2 calibration. Different
colors of cross ponits indicate different ionization parameter $q$: red, log($q$)\textless7.2; green, 7.2\textless log($q$)
\textless7.7; blue, log($q$)\textgreater7.7. Error bar shows the typical error of the measurements. }
\label{meta}
\end{figure}

The results of three calibrations, electron temperatures $T\oiii$, electron densities $n_{\rm{e}}$\ and 
ionization parameter $q$\ are listed in Table\,\ref{metadata}.
To test the reliability of the oxygen abundance, we compare the three calibrations in Figure\,\ref{meta}.
In Figure\,\ref{meta} we plot O3N2 calibrations and \te\ calibrations vs KK04 calibrations, with
different colors indicating different ionization parameter (see Figure\,\ref{meta} for details).
As shown in this figure, good linear correlations are seen with systematic offsets as whole. 
O3N2 calibration is systematically $\sim$0.4 dex lower than KK04 calibration, 
and it flattens when KK04 calibration is lower than 8.5 due to the limited range of O3N2 index. 
\te\ calibrations agree well with O3N2 calibrations if we ignore the values near the lower limit of O3N2 calibration.
The systematic biases of different methods are expected, which are investigated by \citet{kew08} in detail.  
The systemic 0.4 dex bias could be caused by temperature fluctuations or gradients within 
high-metallicity \hii\ regions. In the presence of temperature fluctuations 
or gradients, \oiii\ is emitted predominantly in high-temperature
zones where O$^{2+}$ is present only in small amounts. 
Thus the high electron temperatures estimated from the $\oiii\lambda4363$\ 
line do not reflect the true electron temperature in the \hii\ region, leading to a systematic lower  
estimation of metallicity by as much as 0.4 dex \citep{sta02, sta05, kew08}.
Moreover, the scattering between O3N2 calibrations and KK04 calibrations is not random. It
 correlates with ionization parameter $q$: samples with lower log($q$) tend to have larger measured metallicity. 
Considering all the discussion above, we use KK04 calibrations in the following analysis.

\section{Results}

\subsection{Velocity field}

\begin{figure} \centering
\includegraphics[width=0.45\columnwidth]{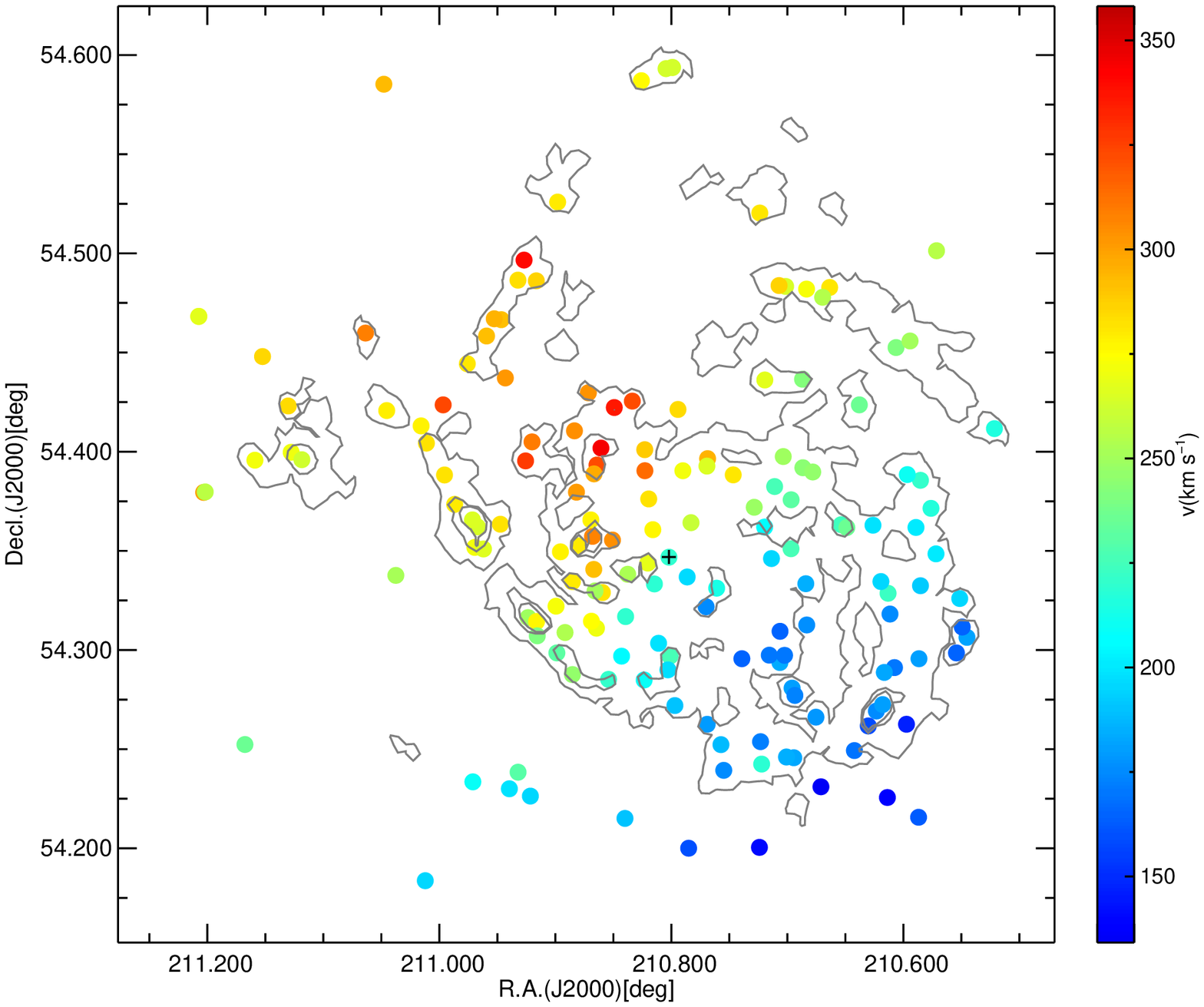}
\includegraphics[width=0.45\columnwidth]{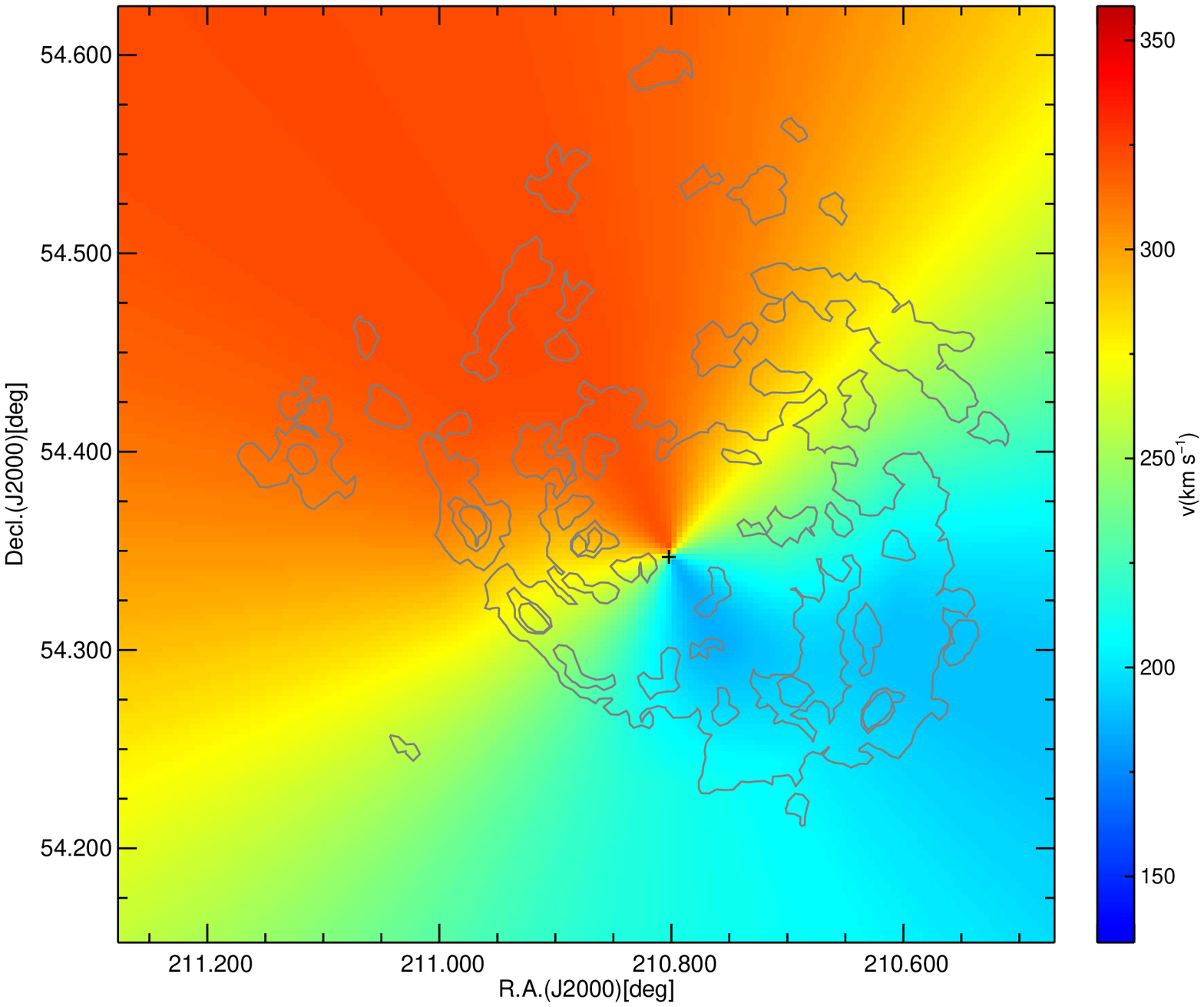}
\caption{Velocity field of M101. Left: the line of sight velocities of \hii\ regions; Right: modeled velocity map using the 
method from \citet{kra06}. Contours show the shape of M101 in \ha\ image, and the center of the galaxy is 
marked with a black cross.}
\label{2d}
\end{figure}

M101 is a giant and nearly face-on spiral galaxy \citep{san74, bosma81}.
The velocity field of M101 have been studied based on the kinematics of HI gas \citep{bosma81} and H$\alpha$-data 
\citep{com79}. In this paper, the \hii\ regions cover a wide range of regions on M101, which enables us to acquire
the velocity map for M101.  
The left panel of Figure\,\ref{2d} shows the line of sight velocities of \hii\ regions in M101, which are good tracers 
of gas movements. The velocities are measured using the SAO xcsao package with IRAF.  

As seen in this panel, a rotating disk is clearly shown.  We then apply the harmonic decomposition method \citep{kra06}
 to model the velocity field.  This method has been widely used to analyze the gas and stellar kinematics in both 
 late-type \citep{ fra94, wong04} and early-type galaxies \citep{ems07, kra08}.
 This method is able to decompose the line of sight velocity map into a series of Fourier components,
  which are kinematic components with different azimuthal symmetry. 
In this work, we broadly expand the Fourier terms to the third order:
\begin{equation}
V(a,\psi)=A_0 (a)+\sum_{n=1}^3 A_n(a) \sin{n\psi} + B_n (a) \cos(n\psi)
\end{equation}
where $\psi$\ is the eccentric anomaly, and $a$ is the length of the
semimajor axis of the elliptical ring. Using this model, we obtained the kinematic
position angle of 36 degree, the inclination of 26 degree, the systemic velocity of 242 km/s and
 the line of sight rotation velocity of 71 km/s, corresponding to a maximum rotation velocity of 168 km/s. 
 These values agree well with the previous modeling of velocity field from \cite{com79}, who found the 
 position angle of 35 degree, the inclination of 27 degree and the systemic velocity of 238 km/s.  
 However, the inclination measured in this work is slightly larger than the result from \cite{bosma81}, who found
 a inclination of 18 degree based on the HI velocity map. This discrepancy may due to the larger HI disks than 
 optical disk and interactions with close companions in outskirts \citep{bosma81, wall97, mihos12}.  
 The modeled velocity map is shown in the right panel of Figure\,\ref{2d} with the same color-coding of the left panel.

\subsection{Stellar population ages}

The light-weighted mean stellar age is a good tracer to the star formation history. Low stellar
age indicates the young stellar population and recent strong star formation activities in the galaxy, 
while high stellar age means old stellar populations with few recent star formation activities. 
The equivalent width (EW) of \ha\ emission line is sensitive to the ratio of present to past star formation rate,
therefore it is expected to be correlated with stellar age, with the higher values the younger populations.

\begin{figure} \centering
\includegraphics[width=0.45\columnwidth,angle=270]{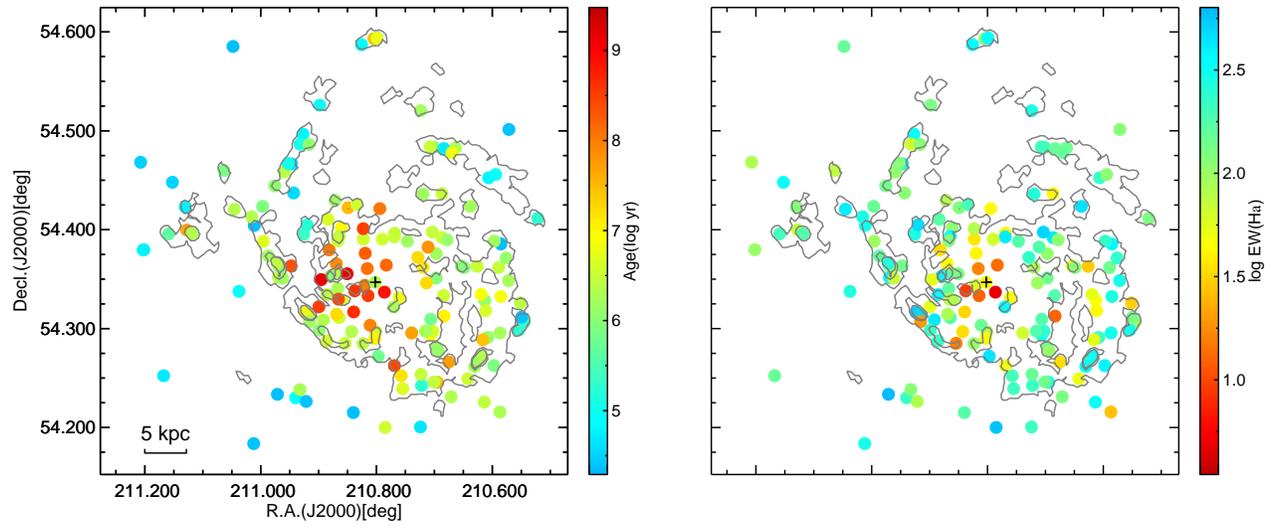}
\caption{The two dimensional distrbution of stellar population age and EW(\ha). Contours show the 
shape of M101 in \ha\ image, and center of the galaxy is marked with a black cross.}
\label{age}
\end{figure}

From the MF-ICA fits, we have estimated the stellar population age by computing light-weighed  
mean age of different stellar population components. The left panel of Figure\,\ref{age} shows the 
two dimensional distribution of the mean stellar age. As shown in this figure, the stellar ages of 
the inner region are generally older than the outskirts in M101. We also present the EW(\ha) of each \hii\ region
in the right panel of Figure\,\ref{age}. The results in both of panels are broadly consistent,
 with older stellar population in the inner regions and younger stellar population in outer regions. 
 This negative stellar age gradient in M101 agrees with the ``inside-out'' disk growth scenario 
 \citep{per13, bez09, tac15}. 

 We note that it appears to have a younger stellar population component at the very center of the bulge (see 
 Figure\,\ref{age}). 
This is also discovered in \citet{lin13} by using SED modelings of resolved multi-band photometric images
from ultraviolet and optical to infrared of M101. They found a resolved bar at the center of M101, and proposed 
 that the bulge of M101 is a disk-like pseudobulge, likely induced by secular evolution of the small bar
 \citep{haw86, ho97, wang12, cole14, lin17}. 


\subsection{Radial Abundance Gradient}

The measurements of radial abundance gradient from \hii\ regions are firstly carried out by 
\citet{sea71} and \citet{smith75}, after that series of works showed up 
\citep{mc85,kenn96,bre07,lin13,san12a,san14,san16b,san17,pil14,bel17}.

\begin{figure} \centering
\includegraphics[width=0.45\columnwidth]{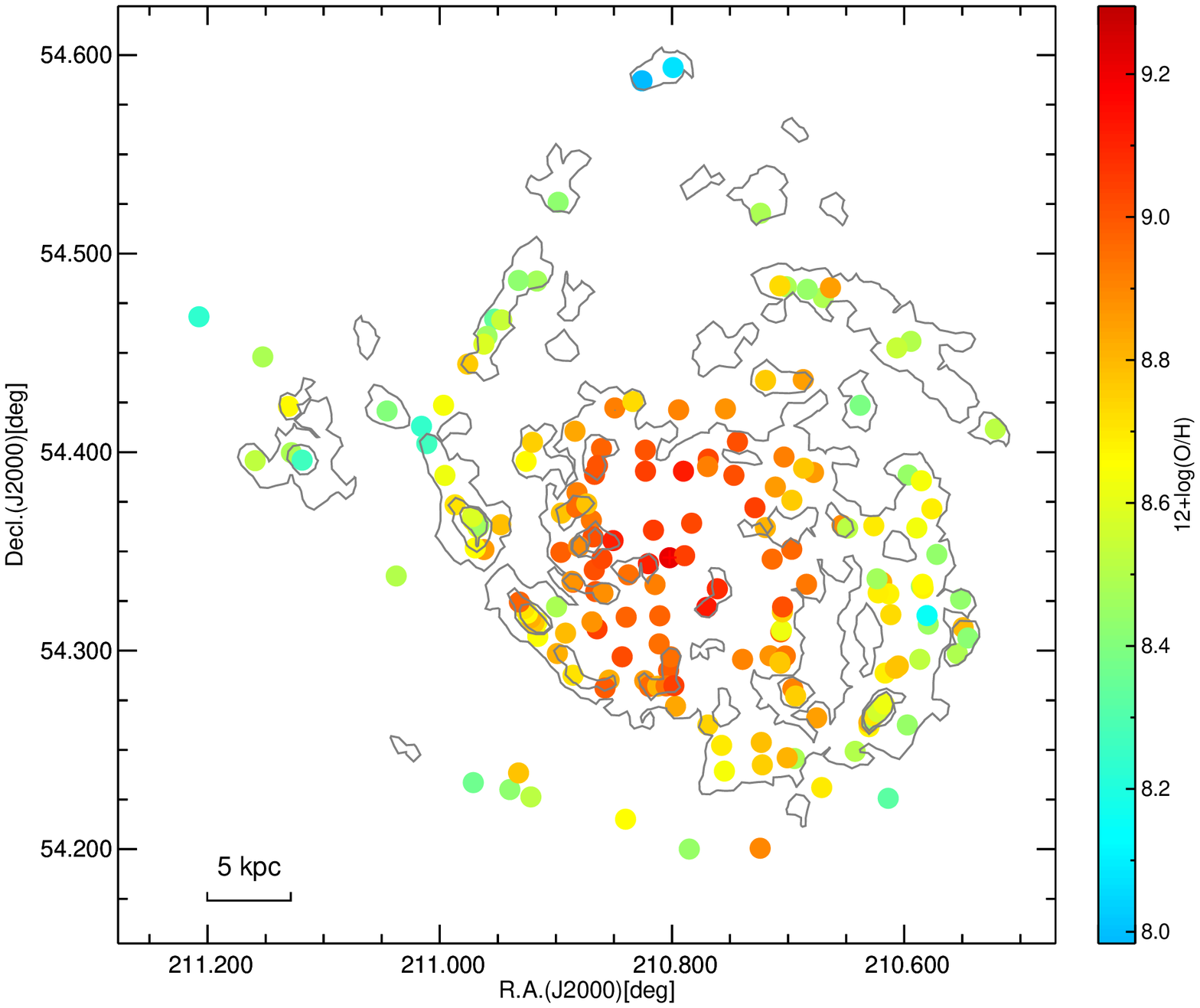}
\includegraphics[width=0.45\columnwidth]{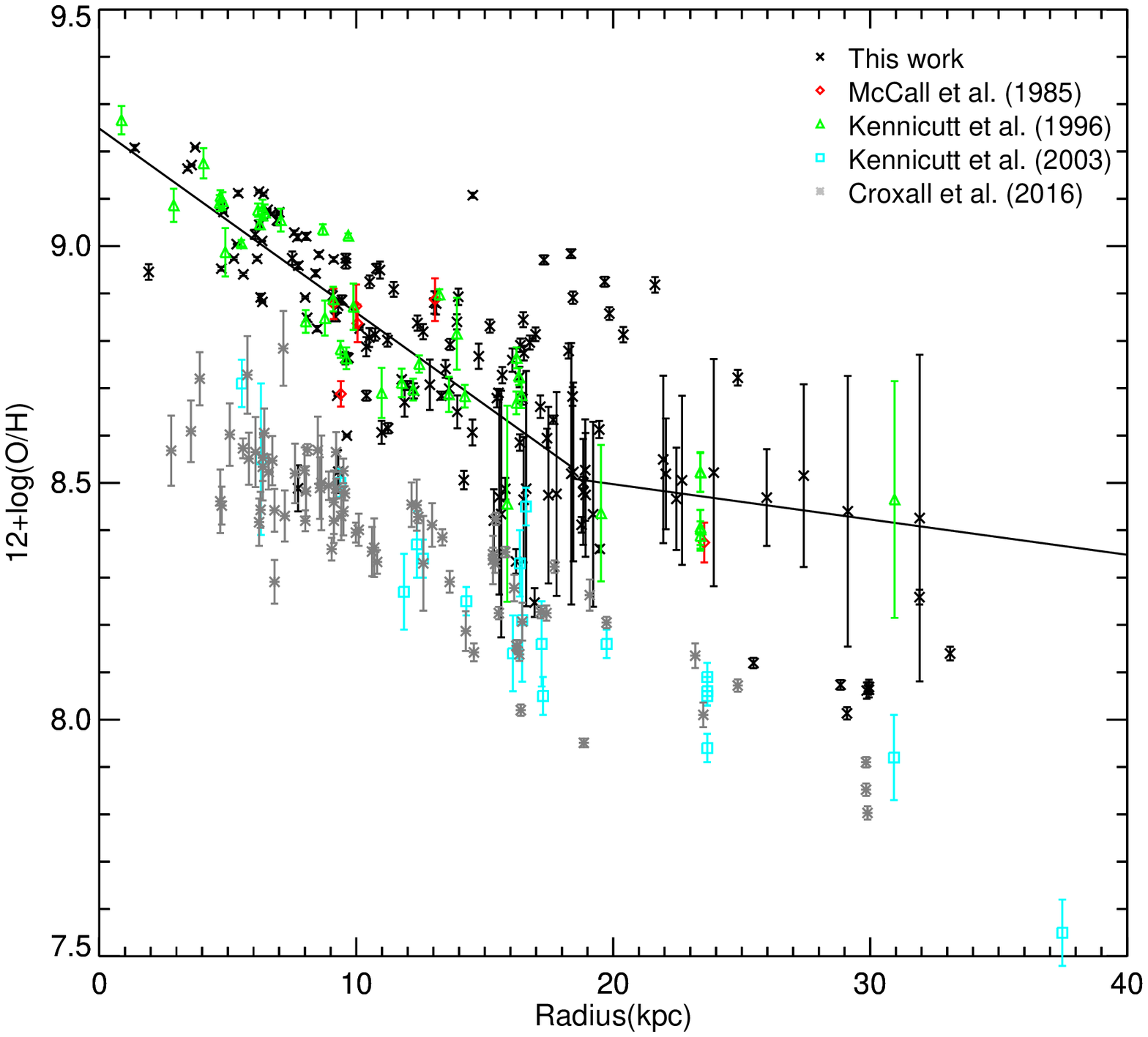}
 \caption{Left: distributions of metallicities of \hii\ regions in M101. Right: Radial metallicity gradient for \hii\ regions. 
Metallicities are determined from KKO4 calibration for this plot.}
\label{gradient}
\end{figure}

There are 173 out of 188 \hii\ regions whose oxygen abundances are computed by the KK04 calibration. 
The distribution of oxygen abundance (left panel) and its radial gradient (right panel) are
 presented in Figure\,\ref{gradient}. Some other works are also shown in different symbols. 
 As the figure shows, the oxygen abundances from the KK04 calibration are 0.5 dex higher than
  \citet{bre07} and \citet{kenn03}, because their calibrations are based on the $T_e$ method. 
However, the metallicity gradients are consistent based on different calibrations and works. 
In addition, our result also shows a break in the gradient at around 18 kpc radius,
so we use two linear least-squares fits:
\begin{equation}\centering
12+\log(\rm{O/H})=9.24-0.0364\times\rm{R}
\end{equation}
for R\textless18 kpc, and
\begin{equation}\centering
12+\log(\rm{O/H})=8.64-0.00686\times\rm{R}
\end{equation}
for R\textgreater18 kpc. The gradient of the inner regions is $-$0.0364 dex kpc$^{-1}$, which
 is slightly steeper than the results from \citet{bre07}, \citet{li13} and \citet{pil14}($-$0.0278, 
$-$0.0268 and $-$0.0293 dex kpc$^{-1}$). 
However, the metallicity gradient becomes flat at the radius greater than 20 kpc. 
To compare with the characteristic gradient provided by 
\citet{ho15}($-$0.39$\pm$0.18 dex R$_{25}^{-1}$), we convert the derived gradient to same scale. For 
inner regions, the gradient is $-$1.18$\pm$0.017 dex R$_{25}^{-1}$, which is steeper 
than characteristic gradients. However, the gradient of the outer region is 
$-$0.221$\pm$0.038 dex R$_{25}^{-1}$, which is flatter.

\citet{ros12} found that gas metallicity increases with stellar mass surface density, which plays a
role in determining the mass-metallicity relation and radial metallicity gradients in spirals. This 
has been confirmed by several researches (e.g.,\citealt{san13}, \citealt{bar16}). Chemical evolution models 
with most important parameters varying smoothly with radius can easily reproduce negative gradients, 
which show that the maxima of both neutral and molecular gas move from the center to 
outskirts through the disk as the galaxy evolves. Thus the star formation is strong in the central 
regions at early times and spread outward through disks as the gas is efficiently consumed. This shapes
present galaxies with high metallicity, low specific star formation rate in the center and negative 
radial metallicity gradients that flatten with time. This is knew as the ``inside-out'' disk growth.

Several studies have found that metallicity gradients flatten to a constant value beyond the isophotal 
radii R$_{25}$ or 2R$_e$\citep{ros11,mar12,san12b,san14,lop15,san16b}.
The mechanism of this gradient flattening is under debate.
One possible mechanism is a slow radial dependence of the star formation efficiency across
over a large galactocentric distances \citep{bre12,est13}. Cosmological simulations that introduce a 
balance between outflows and inflows with the intergalactic medium may also contribute to shaping the 
metal content \citep{opp08,opp10,dav11,dav12}.  
Interactions with satellite galaxies are suggested to enhance the metallicity in the outer 
regions of galaxies, which may also flatten the metallicity gradient \citep{bird12,lop15}.
The real senario is likely to be the mixture of several mechanisms.
We note that the turning point of 
the gradient in M101 is at $\sim$\ 18 kpc, which is much smaller than R$_{25}$($\sim$ 31 kpc).
An alternative explanation is proposed \citet{lep11}, that
the break in the gradient is the ``pumping out'' effect of corotation, which
produces gas flows in opposite directions on the two sides of the radius. The corotation radius of
M101 is 16.7 kpc \citep{scar13}, which is very close to the break radius we find in this work.  
This suggests that the radial break of metallicity in M101 probably caused by the result of corotation.

There are 7 outliers far below the fit line with oxygen abundance lower than 8.2. These \hii\ regions are 
located to the east of M101. 
M101 is not the isolated galaxy, and two close companions (NGC5474 in the east and NGC5477
 in the southeast) are found.
 \citet{kew06} presented some evidence that galaxy pairs have systematically lower oxygen abundance than field 
galaxies, and mergers may also flatten the oxygen abundance gradient. More recently, \citet{san14} found 
that the metallicity gradients are independent of morphology, incidence of bars, absolute magnitude 
and mass, and that the only clear correlation is between merger stage and metallicity gradient, where 
merger progresses flatten the slope of gradients \citep[see also][]{torr12, rich12, kew10, rup10} . M101 might 
be interacting with NGC5474 and NGC 5477, as suggested by 
\citet{wall97, mihos12,mihos13}, who analyzed deep neutral hydrogen observation and deep optical images 
and concluded that M101 is in an ongoing interactions with lower mass companions. These interactions
may dilute the oxygen abundance of these \hii\ regions in gas mixing processes, making the 7 outliers
relatively meta-poor.


\section{Summary}

We have investigated the distributions of physical properties of \hii\ regions in nearby galaxy M101 using 
spectra obtained from the 6.5 m MMT telescope and the NAOC 2.16 m telescope during 2012 to 2014. 
We present the observations, data reductions, and measurements of emission lines. This
is the largest spectroscopic sample of \hii\ regions in M101 by now. 

We estimated the stellar population ages by fit the continuum with MF-ICA approach, 
and the stellar age profile shows an older stellar population in the inner regions
 and a younger in the outer regions.
There is a younger component at the center of the galaxy, which indicates that a recent 
star formation occurred at the center. This might be caused by gas falling into the center
due to the nonaxisymmetric gravity of the bar.

We calculated the electron temperatures $T\oiii$, electron densities $n_e$, and measured oxygen abundances 
with three methods. O3N2 calibrations are consistent with direct \te\ calibrations when ignore the limitation effect
of O3N2 calibration, and both calibrations are systemically $\sim$0.4 dex lower than KK04 calibration, due to temperature fluctuations or gradients within high-metallicity \hii\ regions. O3N2 calibrations
do not consider ionization parameters, which may cause systemic errors.

The oxygen abundance profile shows a negative gradient with a clear break at 20 kpc radius. The gradient is
$-$0.0364 dex kpc$^{-1}$ in the inner region and $-$0.00686 dex kpc$^{-1}$ in the outer region. This is
likely to be the mixture of several mechanisms, such as radial dependence of SFE over large a large galactocentric 
distances, a balance between outflows and inflows with the intergalactic medium and interactions with satellite galaxies. Since the break radius is comparable to the corotation radius (16.7 kpc), the break in the gradient could be the 
``pumping out'' effect of corotation. If this is the case, gas flows play an important role in chemical evolutions 
of galaxies and should not be regarded as minor effects in evolution models. There are several
\hii\ regions with oxygen abundance lower than 8.2, which locate far from major distribution, and we find they 
 are all located in the east of M101, close to its companions, NGC5474 and NGC5477. 
 These low metallicities may be the result of interactions with close companions. 

Both stellar population age distribution and oxygen abundance gradient support the ``inside-out'' disk 
growth scenario: the early gas infall or collapse made the inner region more metal-rich and older,
while the outer disk is enriched more slowly and younger. 
 In order to learn more about the formation and evolution of nearby galaxies, 
 we intend to combine the spectra of \hii\ regions with multi-wavelength
photometry data (especially UV to IR) to investigate the nature of dispersions of IRX-$\beta$\ relation.
Due to the relatively simple star formation history and structure of \hii\ regions,
it would be helpful to find a second parameter to reduce the scatter of IRX-$\beta$\ relation.
This will help us understand further in the interaction between dust and interstellar radiation field, 
as well as the formation and evolution of nearby galaxies.

\section*{Acknowledgments} 

This work uses data obtained through the Telescope Access Program (TAP), which is funded by the 
National Astronomical Observatories of China, the Chinese Academy of Sciences (the Strategic Priority 
Research Program, ``The Emergence of Cosmological Structures'' Grant No. XDB09000000), and the Special 
Fund for Astronomy from the Ministry of Finance. This work 
is supported by the National Basic Research Program of China (973 Program)(2015CB857004), the 
National Natural Science Foundation of China (NSFC, Nos. 11320101002, 11421303, and 11433005), and 
the Youth Innovation Fund by University of Science and Technology of China (No. WK2030220019).

\vspace{5mm}
\facilities{MMT, Beijing:2.16m}
\software{SExtractor \citep{ber96}, xfitfibs, HSRED, MF-ICA \citep{hu16}, MPFIT \citep{mark09}, PyNeb \citep{lur15}}

\clearpage

\begin{deluxetable}{cccrrrrrrrrrrrrrrr} \tablewidth{0pt}
\tabletypesize{\scriptsize}
\rotate
\tablecaption{ Coordinates and Dereddened Line Fluxes of \hii\ Regions in M101}
\tablehead{ ID & R.A. & Decl.& \oii&\neiii&\hd& \hg& \oiii& \oiii& \oiii& \hei& \oi& \nii& \ha& \nii& \sii& \sii& \hb\\ 
 & & &$\lambda$3727&$\lambda$3869&$\lambda$4102&$\lambda$4340&$\lambda$4363&$\lambda$4959&$\lambda$5007&$\lambda$5876&$\lambda$6300&$\lambda$6548&$\lambda$6563&$\lambda$6583&$\lambda$6717&$\lambda$6731&$\lambda$4861}
\startdata
1& 14:03:01.15&54:14:27.0& 276.4&  19.5&  26.5&  48.1&  82.44&  99.4& 294.7&  10.9&   6.5&  9.9& 285.9& 28.4& 22.4& 16.5&  3611.51\\
      &   && 0.6& 0.2& 0.1& 0.3&   0.09& 0.01& 0.8& 0.03& 0.04& 0.07&  1.1& 0.1&   0.11&   0.09&    12.85\\
2& 14:02:44.66&54:23:36.1& 344.0&...&  30.2&  51.6&...&  43.7& 128.5&   8.9&...& 15.6& 228.7& 44.6& 20.3& 14.3&  3478.57\\
      &   && 0.2&...& 0.1& 0.2&...& 0.09& 0.2& 0.1&...& 0.06&  0.6& 0.1&   0.11&   0.09&     7.04\\
3& 14:02:49.49&54:17:42.5& 294.0&...&  28.3&  48.1&...&  49.7& 148.8&  11.3&...& 19.0& 286.0& 54.4& 35.0& 24.1&   276.81\\
      &   && 1.9&...& 0.6& 0.4&...& 0.05& 1.3& 0.4&...& 0.01&  2.4& 0.4&   0.66&   0.32&     2.37\\
4& 14:02:20.22&54:23:12.8& 434.4&...&  40.7&  46.6&...&  51.2& 152.7&  11.9&...& 11.1& 286.9& 31.9& 41.5& 28.1&   131.22\\
      &   && 5.0&...& 1.2& 3.5&...& 0.07& 1.2& 0.6&...& 0.02&  1.9& 0.9&   0.95&   0.92&     2.77\\
5& 14:02:20.65&54:17:48.9& 389.9&  27.1&  26.2&  46.4&  23.73&  94.7& 283.2&  10.3&   8.1&  8.9& 275.3& 25.6& 29.3& 21.3&   510.45\\
      &   && 2.4& 0.9& 0.5& 0.4&   0.35& 0.01& 1.2& 0.3& 0.2& 0.05&  1.3& 0.2&   0.07&   0.18&     6.79\\
6& 14:03:45.54&54:13:53.3& 482.9&  43.4&...&  77.8&...& 103.1& 304.0&...&...&  4.8& 236.9& 13.6& 16.9& 16.6&   174.60\\
      &   && 3.3& 1.3&...& 1.8&...& 0.09& 1.0&...&...& 0.01&  0.8& 0.4&   0.51&   0.24&     1.69\\
7& 14:03:41.73&54:19: 4.0& 374.0&  21.2&  36.3&  56.6&...&  94.3& 275.3&   9.6&   3.9& 10.5& 286.0& 29.9& 14.1& 11.2& 23703.82\\
      &   && 0.1& 0.01& 0.02& 0.1&...& 0.04& 0.2& 0.01& 0.1& 0.02&  0.4& 0.02&   0.03&   0.02&    24.11\\
8& 14:03:34.05&54:18:37.1& 174.1&  12.1&  30.0&  50.7&...&  86.3& 256.0&  12.6&   2.6& 11.9& 252.0& 34.0&  9.1&  7.5& 18061.50\\
      &   && 0.3& 0.1& 0.2& 0.3&...& 0.03& 0.6& 0.1& 0.1& 0.01&  1.2& 0.1&   0.04&   0.05&    63.28\\
9& 14:02:46.64&54:14:49.7& 334.0&  40.9&  29.5&  56.4&   7.29& 125.1& 374.6&   8.6&  12.7& 15.9& 286.0& 45.5& 44.6& 32.2&   354.90\\
      &   && 2.2& 0.9& 0.8& 0.4&   0.38& 0.02& 2.4& 0.2& 0.5& 0.01&  2.2& 0.5&   0.50&   0.27&     2.75\\
10& 14:03:27.54&54:18:45.0& 153.5&...&  31.9&  52.3&...&   7.0&  20.8&   8.6&   3.3& 28.8& 269.0& 82.6& 43.5& 28.3&   442.82\\
      &   && 1.2&...& 0.5& 0.5&...& 0.01& 0.4& 0.2& 0.2& 0.01&  0.9& 0.5&   0.51&   0.25&     1.80\\
11& 14:02:29.51&54:16:14.1& 295.0&  15.8&  25.6&  46.4&...&  92.9& 275.6&  10.9&   6.6& 10.7& 251.7& 30.7& 26.3& 18.0&  1471.36\\
      &   && 0.8& 0.2& 0.3& 0.4&...& 0.01& 0.5& 0.1& 0.2& 0.02&  1.0& 0.2&   0.12&   0.15&     6.10\\
12& 14:02:46.97&54:16:56.1& 280.2&...&  33.0&  36.7&...&  21.6&  64.7&...&   8.8& 26.1& 286.0& 74.9& 40.1& 23.5&   110.45\\
      &   && 4.0&...& 2.1& 1.3&...& 0.02& 1.9&...& 0.7& 0.01&  3.0& 0.7&   0.95&   0.62&     1.72\\
13& 14:03:12.28&54:17:54.2& 222.8&...&  14.0&  49.0&...&  21.2&  63.6&  11.6&...& 27.8& 279.3& 79.8& 44.1& 30.3&   164.75\\
      &   && 2.8&...& 0.8& 2.2&...& 0.01& 1.6& 0.6&...& 0.03&  1.6& 0.8&   1.06&   0.22&     2.31\\
14& 14:02:20.29&54:20: 1.4& 275.7&  21.6&  28.6&  47.3&...& 100.1& 300.4&  11.0&   6.2&  7.0& 286.2& 20.2& 18.6& 13.5&   576.54\\
      &   && 2.5& 0.6& 0.8& 0.8&...& 0.02& 1.7& 0.2& 0.2& 0.04&  2.1& 0.2&   0.17&   0.22&     5.98\\
15& 14:02:12.15&54:19:37.9& 175.2&  78.7&...&...&...& 227.3& 681.8&...&...&...& 237.9&...&...&...&    34.70\\
      &   &&11.6& 6.4&...&...&...& 0.04& 6.5&...&...&...&  2.3&...&...&...&     1.69\\
\enddata
\tablecomments{\ Line fluxes are normalized to \hb line, after correcting for reddening. \hb$\lambda$4861\ is the measured 
\hb\ flux in units of $\times10^{-17}$erg s$^{-1}$\ cm$^{-2}$, corrected for extinction.}
\end{deluxetable}
\label{data}
\clearpage

\begin{deluxetable}{ccccccc} \tablewidth{0pt}
\tabletypesize{\scriptsize}
\tablecaption{Gas Conditions and Oxygen Abundance Measurements}
\tablehead{ ID & $T\oiii$ & $n_e$ & $\log q$ &\multicolumn{3}{c}{\zoh}\\
 \cline{5-7}   &  (K)   & (cm $^{-3}$)& $(\log$ cm s$^{-1})$& O3N2 & KK04 & $T_e$ }  
\startdata
1&   10120 $\pm$     1134&      76 $\pm$        12& 7.77 $\pm$ 0.153& 8.27 $\pm$ 0.023& 8.68 $\pm$ 0.009& 8.32 $\pm$ 0.933 \\
2&... &      29 $\pm$        7& 7.47 $\pm$ 0.019& 8.46 $\pm$ 0.013& 8.83 $\pm$ 0.002&...  \\
3&... &... & 7.60 $\pm$ 0.654& 8.45 $\pm$ 0.070& 8.85 $\pm$ 0.009&...  \\
4&... &... & 7.37 $\pm$ 1.145& 8.37 $\pm$ 0.096& 8.68 $\pm$ 0.012&...  \\
5&   13696 $\pm$     1415&      62 $\pm$        11& 7.57 $\pm$ 0.388& 8.25 $\pm$ 0.036& 8.59 $\pm$ 0.017& 7.94 $\pm$ 0.820 \\
6&... &     539 $\pm$       30& 7.55 $\pm$ 3.646& 8.17 $\pm$ 0.036& 8.63 $\pm$ 0.009&...  \\
7&... &     170 $\pm$        29& 7.58 $\pm$ 0.056& 8.31 $\pm$ 0.006& 8.60 $\pm$ 0.002&...  \\
8&... &     215 $\pm$        37& 8.05 $\pm$ 0.394& 8.32 $\pm$ 0.021& 8.85 $\pm$ 0.004&...  \\
9&    9519 $\pm$     2096&      53 $\pm$        20& 7.68 $\pm$ 0.343& 8.29 $\pm$ 0.057& 8.51 $\pm$ 0.020& 8.52 $\pm$ 1.876 \\
10&... &... & 7.33 $\pm$ 0.773& 8.78 $\pm$ 0.078& 9.11 $\pm$ 0.003&...  \\
11&... &... & 7.72 $\pm$ 0.283& 8.29 $\pm$ 0.017& 8.69 $\pm$ 0.008&...  \\
12&... &... & 7.39 $\pm$ 1.443& 8.61 $\pm$ 0.165& 8.95 $\pm$ 0.010&...  \\
13&... &... & 7.53 $\pm$ 1.244& 8.62 $\pm$ 0.147& 9.02 $\pm$ 0.010&...  \\
14&... &      62 $\pm$        15& 7.65 $\pm$ 0.572& 8.21 $\pm$ 0.050& 8.47 $\pm$ 0.206&...  \\
16&... &      59 $\pm$        16&... &... & 9.16 $\pm$ 0.004&...  \\
17&... &... & 7.96 $\pm$ 1.300& 8.45 $\pm$ 0.039& 8.89 $\pm$ 0.008&...  \\
18&   15259 $\pm$      764&      69 $\pm$        27& 7.65 $\pm$ 0.539& 8.20 $\pm$ 0.043& 8.46 $\pm$ 0.133& 7.79 $\pm$ 0.390 \\
19&... &      20 $\pm$        7&... &... & 9.03 $\pm$ 0.004&...  \\
20&... &... & 7.44 $\pm$ 2.586& 8.57 $\pm$ 0.114& 8.95 $\pm$ 0.018&...  \\
21&... &     113 $\pm$        34& 7.48 $\pm$ 0.211& 8.68 $\pm$ 0.023& 9.07 $\pm$ 0.002&...  \\
22&... &... & 7.76 $\pm$ 2.431& 8.25 $\pm$ 0.088& 8.61 $\pm$ 0.018&...  \\
23&... &     164 $\pm$        35&... &... &......&...  \\
24&... &      25 $\pm$        9& 7.42 $\pm$ 0.569& 8.59 $\pm$ 0.115& 8.96 $\pm$ 0.011&...  \\
25&... &... & 7.69 $\pm$ 2.493& 8.55 $\pm$ 0.101& 9.02 $\pm$ 0.006&...  \\
26&... &      55 $\pm$        12& 7.43 $\pm$ 0.433& 8.44 $\pm$ 0.038& 8.68 $\pm$ 0.010&...  \\
27&... &... &... &... & 8.86 $\pm$ 0.013&...  \\
28&... &     498 $\pm$       22& 7.42 $\pm$ 0.453& 8.29 $\pm$ 0.089& 8.66 $\pm$ 0.024&...  \\
29&... &     923 $\pm$       83& 7.39 $\pm$ 0.445& 8.30 $\pm$ 0.255& 8.51 $\pm$ 0.179&...  \\
30&... &     182 $\pm$       15& 7.56 $\pm$ 3.270& 8.33 $\pm$ 0.144& 8.79 $\pm$ 0.014&...  \\
\enddata
\end{deluxetable}
\label{metadata}
\clearpage

\end{document}